\documentclass[review,11pt]{elsarticle}

\usepackage{graphicx}
\usepackage{bm}
\usepackage{amssymb,amsfonts,amsmath}
\usepackage[pdftex]{color}
\usepackage{float}
\usepackage{epsfig}
\usepackage{booktabs} 
\usepackage{array} 
\usepackage{paralist}
\usepackage[english]{babel}
\usepackage[normalem]{ulem}
\graphicspath{{figs/}}
\usepackage{multirow}
\usepackage{soul}

\usepackage{lineno}

\newcommand{\0} {\textbf{0}}
\newcommand{\Lin} {\mathbb{L}\mathtt{in}}

\newcommand{\Skw}{{\mathbb{S}\textrm{kw}}}

\newcommand{\Rot} {\mathbb{R}\mathtt{ot}}

\newcommand{\skw} {\mathrm{skw}\,}

\newcommand{\sym} {\mathrm{sym}\,}

\newcommand {\Fe} {\mathbf{F}_e}
\newcommand {\Fv} {\mathbf{F}_v}

\newcommand {\Bc}  {\mathcal{B}}
\newcommand {\Br}  {\mathcal{B}_r}

\newcommand {\Ec}  {\mathcal{E}}

\newcommand {\Gc}  {\mathcal{G}}

\newcommand {\Pc}  {\mathcal{P}}

\newcommand {\Tc}  {\mathcal{T}}
\newcommand {\Uc}  {\mathcal{U}}
\newcommand {\Vc}  {\mathcal{V}}
\newcommand {\Wc}  {\mathcal{W}}
\newcommand {\ab} {\mathbf{a}}

\newcommand {\eb} {\mathbf{e}}

\newcommand {\nb} {\mathbf{n}}

\newcommand {\ub} {\mathbf{u}}
\newcommand {\vb} {\mathbf{v}}

\newcommand {\wb} {\mathbf{w}}
\newcommand {\zb} {\mathbf{z}}
\newcommand {\Ab} {\mathbf{A}}
\newcommand {\Bb} {\mathbf{B}}
\newcommand {\Cb} {\mathbf{C}}
\newcommand {\Db} {\mathbf{D}}
\newcommand {\Eb} {\mathbf{E}}
\newcommand {\Fb} {\mathbf{F}}
\newcommand {\Gb} {\mathbf{G}}

\newcommand {\Ib} {\mathbf{I}}
\newcommand {\Lb} {\mathbf{L}}

\newcommand {\Qb} {\mathbf{Q}}
\newcommand {\Rb} {\mathbf{R}}
\newcommand {\Sb} {\mathbf{S}}
\newcommand {\Tb} {\mathbf{T}}

\newcommand {\Vb} {\mathbf{V}}

\newcommand {\Wb} {\mathbf{W}}

\newcommand {\Co} {\mathbb{C}}

\newcommand{\Grad} {\mathrm{Grad}\,}
\newcommand{\Div} {\mathrm{Div}\,}

\newcommand{\asm}{\text{\bfseries\slshape a\/}}

\newcommand{\Asm}{\text{\bfseries\slshape A\/}}
\newcommand{\Aref}{\boldsymbol{\mathsf{A}}_0}

\newcommand{\arem}{\asm}
\newcommand{\aref}{\asm_0}
\newcommand{\av}{\asm_v}
\renewcommand{\ae}{\Fe\av}
\newcommand{\Av}{\Asm_v}
\newcommand{\Dv}{\Db_v}

\newcommand{\Arem}{\Asm}
\newcommand{\Ce}{\Cb_e}
\newcommand{\Be}{\Bb_e}
\newcommand{\Omegab}{\bm\Omega}

\journal{}

\begin{document}

\title{A structurally frame-indifferent model for anisotropic visco-hyperelastic materials}
\author[r1]{J. Ciambella\corref{cor1}}

\author[r1]{P. Nardinocchi}

\cortext[cor1]{Corresponding Author: J. Ciambella, Dipartimento di Ingegneria Strutturale e Geotecnica, via Eudossiana 18, I-00184 Roma, tel: 0039 06 44585293, fax: 0039 06 4884852.}
\address[r1]{Dipartimento di Ingegneria Strutturale e Geotecnica, Sapienza Universit\`a di Roma, via Eudossiana 18, I-00184 Roma, Italy}
\begin{abstract}
One of the main theoretical issues in developing a theory of anisotropic viscoelastic media at finite strains lies in the proper definition of the material symmetry group and its evolution with time.  In this paper the matter is discussed thoroughly and addressed by introducing a novel anisotropic remodelling equation compatible with the principle of structural frame indifference, a requirement that every inelastic theory based on the multiplicative decomposition of the deformation gradient must obey to.
The evolution laws of the dissipative process are 
completely determined by two scalar functions, the elastic strain energy and the dissipation densities. 
The proper choice of the dissipation function allows us to reduce the proposed model to the Ericksen anisotropic fluid, when deformation is sufficiently slow, or to the anisotropic hyperelastic solid for fast deformations. 
Finally, a few prototype examples are discussed to highlight the role of the relaxation times in the constitutive response.
\end{abstract}

\begin{keyword}
continuum mechanics; nonlinear anisotropic elasticity; nonlinear anisotropic viscosity.\end{keyword}

\maketitle


\section{Introduction}
Anisotropic soft solids are a class of materials ubiquitously found either in nature and in artificially made structures. Generally, they are constituted by a (soft) homogeneous matrix with (stiff) reinforcing fibres which equally contribute to their mechanical response. Examples includes biological tissues, such as muscles and arteries \citep{deGennes:1995,Kuhl:2005,Dicarlo:2006}, elastomers \citep{Saxena:2014,Stanier:2016,Ciambella:2017} and soft gels \citep{DeSimone:2007,Sawa:2010,Bosnjak:2019}, to cite but a few. Despite being macroscopically diverse, the materials above share microstructural similarities due to the presence of long--chain molecules intertwined to each other, which form a spaghetti--like bundled structure with a high degree of flexibility. In response to  an externally imposed stress, the long chains may alter their configurations relatively rapidly due to their high mobility. The requirement of linking the chains into a network structure is associated with solid-like features, which allow the material to be stretched up to about ten times of its original length. In addition, the long molecules may partially slide onto each other causing an internal reorganization which, macroscopically, manifests itself in a viscous-like behaviour.  The combination of these two effects allows the materials to exhibit simultaneously the characteristics of a viscous fluid and of an elastic solid.

To date, several modelling strategies have been proposed to describe the viscoelastic behaviour of anisotropic soft materials \cite{Diani:2006,Nguyen:2007,Nedjar:2007,Latorre:2015,Latorre:2016,Balbi:2018,Liu:2019,Bosnjak:2020}. All of them take up the proper definition of the (elastic) long-term material response and the formulation of the evolution laws of the dissipative process accounting for the internal material response. Moreover, they also share the use of the multiplicative decomposition of the deformation gradient \cite{Lee:1969} as a tool to distinguish elastic and viscous deformations, and the consequential consideration of a natural state of the body.
%
 The use of the multiplicative decomposition in the context of anisotropic inelasticity  raises several issues, the most significant one being the definition of the internal material symmetries in the natural state. In this respect, the different approaches used in the literature can be divided in two main classes. One includes those approaches which assume the viscous deformation not to altering the internal material structure, thus the material symmetry group, induced by the reorientation of the reinforcing fibres, remains the same in the reference configuration and in the natural state \cite{Nedjar:2007,Latorre:2015,Latorre:2016,Bosnjak:2020}. The other class includes those models which make the assumption that the symmetry group evolves with the viscous part of the deformation \cite{Nguyen:2007,Liu:2019}. Although the first approach seems adequate in crystal plasticity, where the plastic sliding of the crystalline planes may not modify the material symmetry (see, for instance, \citep{Dashner:1986}), experimental evidence on amorphous polymers suggests that the inelastic part of the deformation must play a role in the evolution of the symmetry group \cite{Boyce:1988,Fares:1991}. In all cases, one must formulate the proper evolution laws of the dissipation process, which are linked to the evolution of the natural state of the body, in terms of the different state variables.\footnote{See \cite{Liu:2019} for a comprehensive discussion which is beyond the scope of the present work.} 
In the context of isotropic viscoelasticity, fo instance, the dissipation inequality was used in \cite{Reese:1998} to define the evolution law of the viscous strain to guarantee that the dissipation be positive for every realizable process. An extension of that framework was presented and discussed in \cite{Latorre:2015,Latorre:2016} to include anisotropic materials. Similarly, in \cite{Nguyen:2007} the dissipation inequality allowed the authors to define the evolution equations by assuming two different viscous deformations and evolution laws for the fibre and the matrix; as such two independent characteristic times were associated to the viscous flow. More recently, the same distinct decomposition of the deformation gradient was implemented in \cite{Liu:2019}, where the evolution laws were written in terms the isotropic and anisotropic viscous components of the symmetric Piola-Kirchhoff stress. Yet, such a model includes only two characteristic times, one for the matrix and one for the fibres. In both \cite{Nguyen:2007} and \cite{Liu:2019}, the evolution of the anisotropy axis is not affected by the viscous part of the deformation and the description of the material anisotropy in the natural state remains is the same as the reference configuration.

All these approaches can be viewed within the unifying theory of material remodelling \cite{Rodriguez:1994,Dicarlo:2002}, with the various internal variables defining the viscoelastic behaviour of the body determined by different evolution laws. 
Within that framework, this work wants to establish a theory of viscoelastic anisotropic bodies based on a new (remodelling) balance equation which delivers, once the proper constitutive information are used, the evolution laws for the dissipation process. The constitutive issues are addressed by invoking three basic principles: the principle of indifference to change in observer, the dissipation principle and the principle of structural-frame indifference \cite{Green:1971}. Whereas the first two are enforced in every mechanical theory, the latter comes out from the multiplicative decomposition of the deformation gradient. It will be shown that structural-frame indifference makes the elastic strain energy and the dissipation function free of the rotational indeterminacy of the natural state \cite{Gurtin:2010}. 
Although the theory developed here can be applied to materials that have complex internal structures, the paper  focus on those materials having a single preferred direction and so are called transversely isotropic. Such a choice maintains the derivation of all equations simple, yet it allows the description of a large class of materials that are of interest in many engineering applications \citep{Warner:2003,Fukunaga:2008,Stanier:2016,Turzi:2016,Ciambella:2017,Ciambella2:2019}. By assuming the material to be transversally isotropic in the natural state, and that the material group changes with the viscous part of the deformation, it is shown that the requirement of material symmetry and of structural frame-indifference coincides. 
The rational structure of the model is easily implemented in a finite element code and, for states close to the thermodynamic equilibrium, allows us to recover by linearisation several known formulations.
Finally, by focusing on a particular form of the dissipation function, three independent relaxation times are considered and it is shown that they are associated to the isotropic contribution of the matrix, and to the additional dissipation introduced by the fibres. Their contribution to the mechanical response is discussed thoroughly through the prototypical example of relaxation under confined uniaxial extension.
%
\section{Kinematics}
\label{Kin}
%
We set the kinematics of the continuum within the framework induced by the Kr\"oner-Lee decomposition of the deformation gradient, which is largely used for modelling inelastic deformation of materials \citep{Eckart:1948, Lee:1969, Green:1971,Galante:2013}.\\
We assume that a region $\Br$ of the three--dimensional  Euclidean space $\Ec$ is the reference configuration of the body and denote with $p :\Br\times \Tc\to\Ec$ the time-dependent map called transplacement which assigns at each point $X\in\Br$ a point $x=p(X,t)$ at any instant $t$ of the time interval $\Tc=(t_0,T)$. We identify as  current configuration $\Bc_t=p(\Br,t)$ the region occupied by the body  at time $t$ and set $p(\Br,t_0)=\Br$. Finally, we introduce the displacement field $\ub:\Bc_r\times \Tc\to\Vc=T\Ec$ such that $x=X +\ub(X,t)$, which is one of the state variable of the problem.

\begin{figure}
\begin{center}
\begin{footnotesize}
\def\svgwidth{.7\textwidth}
   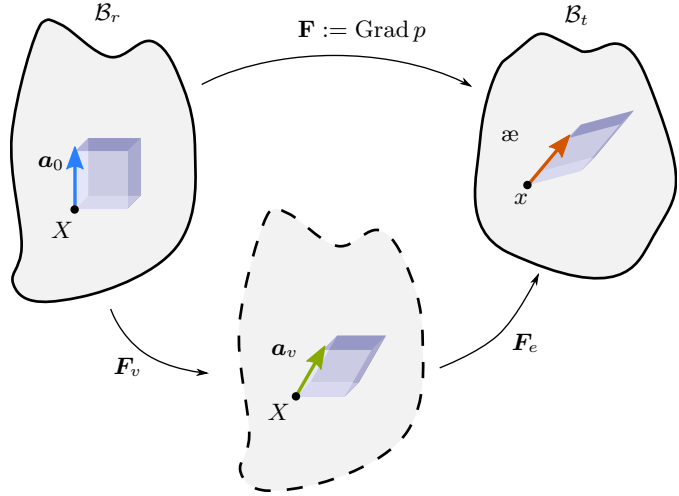
\end{footnotesize}
\end{center}
\label{figure:1}
\caption{K\"roner-Lee decomposition of the deformation gradient $\Fb$ into elastic (reversible) $\Fe$ and viscous (irreversible) $\Fv$ processes. Reference, relaxed and current body elements are shown as parallelepipeds. The material fibre $\ab_0$  (reference fibre) and $\ab_v$ (dragged-by-the-viscous-deformation fibre) at $X$ are represented together with the current fibre $\Fb_e\ab_v$ at $x$. A dashed line is used to sketch the relaxed  state and evidence the fact that the physical pieces that constitute the relaxed state may not fit together.}
\end{figure}

According to the Kr\"oner-Lee decomposition (see Fig.~\ref{figure:1}), the deformation gradient $\Fb:=\Grad p$ is decomposed into viscous  (irreversible) $\Fv$ and elastic (reversible) $\Fe$ deformation tensors\footnote{Despite using the term \emph{deformation} to indicate $\Fe$ and $\Fv$, these tensors may not be the gradient of any maps.}, such that
\begin{equation}
\Fb=\Fe\,\Fv\,.
\label{kronerlee}
\end{equation}
\noindent The viscous deformation $\Fv$ is a smooth tensor-valued field with positive Jacobian determinant: $J_v:=\det\Fv>0\,$. It is the manifestation of the internal material reorganization, which we will call \textit{viscous relaxation}, and is the other state variable of the problem. The tensor $\Fv(X,t)$ acts on a body element at $X\in\Br$ and maps it into its relaxed (or natural) state at time $t$.  As noted above, in general the relaxed state may not be described by a placement, meaning that $\Fv$ may not be the gradient of any map, or in other terms, there is no way to let each body element relaxing to its zero-stress state without removing the surrounding elements \cite{Rodriguez:1994,Dicarlo:2002}. Indeed, it is the elastic reversible deformation $\Fe$ that makes the tensor field $\Fb=\Fe\Fv$ integrable. In the following, we will call $J=\det\Fb$ and so we write $J=J_e\,J_v$ with $J_e=\det\Fb_e$.\footnote{Although polymers are usually assumed incompressible so that $J_e=1$, we do not enforce this condition here analytically, but we will consider almost incompressible material models in the numerical examples.}\\
We must point out that the decomposition \eqref{kronerlee} is not unique and, in fact, an arbitrary local rotation $\Qb\in\Rot$ can be superposed to the viscous deformation $\Fv$ and maintain the natural state unaltered. Namely, if we set
\begin{equation}\label{F+}
\Fe^+=\Fe\Qb^T\quad\textrm{and}\quad
\Fv^+=\Qb\Fv\,,\quad\textrm{then}\quad 
\Fb=\Fe\Fv=\Fe^+\Fv^+
\end{equation}
for any $\Qb\in\Rot$ \citep{Green:1971,Dicarlo:2002}. Following \citep{Green:1971}, we call \emph{structural space} the set of relaxed states at $X\in\Br$ that corresponds to the same current configuration described by $\Fb$. In this sense, Eq.~\eqref{F+} shows that the macroscopic deformation $\Fb$ is insensitive on a rigid body motion superimposed on the structural space, and so it is expected that all the constitutive functions of the model obey to this basic requirement. The invariance property \eqref{F+} was first defined in \citep{Green:1971}, wherein the so-called \emph{structural-frame indifference} was introduced to overcome the consequent lack of uniqueness of the natural state. In Sec.~\ref{SFISec}, we will analyse and discuss the consequences of structural-frame indifference in details.

The internal material structure is described through the unit vector field $\aref:\Bc_r \to\Vc$, which represents the fibre field in the reference configuration, and the associated orientation tensor  $\Aref = \aref\otimes\aref$, sometimes denoted as structural tensor or fabric tensor. We further assume that the viscous deformation $\Fv$ may alter the material structure and so we introduce the corresponding dragged-by-viscous-relaxation orientation tensor as $\Arem_v = \arem_v\otimes\arem_v$ with $\arem_v=\Fb_v\aref/\vert\Fb_v\aref\vert$ (see Fig.~\ref{figure:1}).
Accordingly, the invariance group of the material structure changes from $\Gc_r$, that is the set of all rotations $\Qb\in\Rot$ such that $\Qb \Aref\Qb^T = \Aref$  to $\Gc_v$, that is the set of all rotations such that $\Qb \Ab_v\Qb^T = \Ab_v$. Finally, we call $\ab=\Fb_e\ab_v$  the orientation of the fibre in the current configuration and $\Ab=\Fb_e\Ab_v\Fb_e^T$ the corresponding structural tensor.  
%
%
\subsection{Deformation rates and velocity fields}
%
The time derivative of the deformation gradient $\dot\Fb = d \Fb/d t$ and its elastic $\dot\Fb_e$ and viscous $\dot\Fb_v$ components are related by 
\begin{equation}\label{rates0}
\dot\Fb=\dot\Fb_e\Fb_v + \Fb_e\dot\Fb_v\,,\qquad
\dot\Fb_e=(\Fb\Fb_v^{-1})^\cdot=\Lb\Fb_e-\Fb_e\Lb_v\,,
\end{equation}
where $\Lb:=\Grad\vb=\dot\Fb\Fb^{-1}$ is the velocity gradient (or deformation-rate tensor) and $\Lb_v=\dot\Fb_v\Fb_v^{-1}$ is the viscous deformation-rate tensor. Accordingly, the time derivatives of the elastic $\Cb_e=\Fb_e^T\Fb_e$ and visible $\Cb=\Fb^T\Fb$ left Cauchy-Green strain tensors are 
\begin{equation}\label{rates1}
\dot\Cb_e=(\Fb_e^T\Fb_e)^\cdot = 2\,\Fb_e^T\Db\Fb_e - 2\,\sym(\Cb_e\Lb_v)\quad\textrm{and}\quad
\dot\Cb=2\,\Fb^T\Db\Fb\,,
\end{equation}
where the stretch rate tensor $\Db=\sym\Lb$ has been introduced.\footnote{Throughout the paper $\sym$ and $\skw$ will be used to indicate the symmetric and skew-symmetric part of tensors.}  By defining the viscous stretch rate $\Db_v=\sym\Lb_v$ and the viscous spin $\Wb_v=\skw\Lb_v$, the rate of change of $\av$ is 
\begin{equation}
\dot{\ab}_v = \frac{d}{dt}\left(\frac{\Fv\,\aref}{\vert \Fv\,\aref\vert}\right)=\left( \Wb_v+\Db_v\Av-\Av\Db_v\right)\av\,.
\end{equation} 
Accordingly,
\begin{equation}\label{rates2}
\dot{\Arem}_v=\dot{\arem}_v\otimes\arem_v+\arem_v\otimes\dot{\arem}_v 
=2\,\sym(\Db_v\Arem_v) -2\,\Arem_v\Db_v\Arem_v +[\Wb_v,\Arem_v]\,.
\end{equation}
The operator $[\cdot,\cdot]$ is known as \emph{commutator} and defined by 
\begin{equation}\label{comm}
[\Ab,\Bb]=\Ab\Bb-\Bb\Ab,\quad \forall \Ab,\Bb\in \Lin\,.
\end{equation}
For later use, it is convenient to calculate how velocity fields transform under the invariance requirement~\eqref{F+}. In particular, when $\Fe^+=\Fe\Qb^T$ and $\Fv^+=\Qb\Fv$, one has
%
%
\begin{equation}\label{LvDv}
\Lb_v^+=\Qb\Lb_v\Qb^T+\Omegab\quad\textrm{and}\quad
\Db_v^+=\Qb\Db_v\Qb^T\,, 
\end{equation}
with $\Omegab=\dot{\Qb}\Qb^T\in\Skw$. These two expressions will be used in the next section to introduce suitable simplification in the constitutive functions.
%
\section{Balance equations and thermodynamics}
\label{Bet}
%
The balance equations of the model are derived into two steps from the principle of virtual  working. 
Firstly, we introduce the following continuous, linear, real-valued functionals on the space of actual velocities $(\dot\ub,\Lb_v)$, denoted as external and internal workings, respectively:
\begin{equation}
\Wc_e(\dot\ub)=\int_{\Bc_r}\zb\cdot\dot\ub   + \int_{\partial\Bc_r}\mathbf s\cdot\dot\ub
\quad\textrm{and}\quad\Wc_i(\dot\ub,\Lb_v)=\int_{\Bc_r}(\Sb\cdot\dot\Fb   +  \Gb\cdot\Lb_v)\,.
\label{WorkingRate}
\end{equation}
The external working $\Wc_e(\dot\ub)$ is expended on $\dot\ub$ by the forces per unit (reference) volume and area $\zb$ and $\mathbf s$, respectively, whereas any external actions working-conjugate of $\Lb_v$ are neglected, since we have assumed that viscous remodelling is a purely passive process. The internal working $\Wc_i(\dot\ub,\Lb_v)$ is expended on $\dot\Fb=\Grad{\dot\ub}$ by the Piola--Kirchhoff reference stress  tensor $\Sb$, and we indicate with $\Gb$ the internal action working-conjugate to $\Lb_v$ and call it the \textit{remodelling inner action}. 

Secondly, we derive balance equations and boundary conditions by enforcing the \emph{principle of virtual working}, that is the requirement that, for any given subregion $\Pc\subset\Bc_r$ of the body, the external and internal workings are the same, i.e., $\Wc_e(\wb)=\Wc_i(\wb,\Vb)$ for all virtual velocities $(\wb,\Vb)\in\Vc\times\Lin$ corresponding to the actual velocities $(\dot\ub,\Lb_v)$. In formulae, through a standard derivation, one obtains
\begin{equation}\label{be}
\Div\Sb + \zb =\0\quad\textrm{and}\quad
\Gb=\0\,\,\,\textrm{in}\,\,\Bc_r\,,
\end{equation}
\begin{equation}\label{bcmec}
\ub=\hat\ub\,\,\,\textrm{in}\,\,\partial_u\Bc_r \quad\textrm{and}\quad\Sb\nb=\mathbf{s}\,\,\,\textrm{on}\,\,\partial_t\Bc_r\,,
\end{equation}
where $\partial_u\Bc_r$ and $\partial_t\Bc_r$ are the parts of the boundary $\partial\Bc_r$ where displacements and tractions are prescribed, respectively, and $\nb$ is the unit normal to $\partial_t\Bc_r$. It is worth noting that the balance equation \eqref{be}$_2$ prescribes that the remodelling inner action  must be zero, due to the absence of external actions. Yet this term may be used to introduce suitable multiphysics coupling in the model \citep{Ciambella2:2019}.

Consistency with thermodynamics makes mandatory prescribing a positive dissipation, which is expressed as the difference between the external working $\Wc_e(\dot\ub)$ and the rate of elastic strain energy. With the balance equations \eqref{be}-\eqref{bcmec} on hand and assuming that both dissipation and elastic strain energy can be represented in terms of their (specific) densities $\delta$ and $\varphi$ per unit of mass, one has
\begin{equation}\label{dis}
\int_{\Bc_r} \varrho_r\,\delta = \int_{\Bc_r}(\Sb\cdot\dot\Fb  + \Gb\cdot\Lb_v) -\int_{\Bc_r}\varrho_r\,\dot\varphi\ge 0\,,
\end{equation}
with $\varrho_r$ the reference mass density.\footnote{We further note that $\varrho_r=J_v\varrho_v = J\varrho$, with $\varrho_v$ and $\varrho$ the mass densities per unit relaxed and actual volumes, which implies that during deformation mass is conserved.}\\
In developing the constitutive theory, we shall assume first that the elastic strain energy density $\varphi$ at each $X\in\Bc_r$ depends on the elastic deformation $\Fe$ and on the orientation tensor $\Av$, which is used to convey the information on the internal material structure. 
This modelling choice highlights the differences in our approach from others in the literature in which it is assumed that the internal material structure  is unaltered by $\Fv$ and an elastic strain energy density depending on $\Aref$ is typically assumed.\footnote{Indeed, although this seems adequate in crystal plasticity, where the plastic sliding of the crystalline planes may not modify the material symmetry (see the example in \citep{Dashner:1986}), experimental evidence on amorphous polymers  seems to suggest that the inelastic part of the deformation must play a role in the evolution of the symmetry group \cite{Boyce:1988,Fares:1991}.}

Likewise, the dissipation density $\delta$ is assumed to depend both on the viscous deformation-rate tensor $\Lb_v$ and on $\Av$. As a result, with a little abuse of notations, we write
\begin{equation}
\varphi = \varphi(\Fe,\Av)\qquad \text{and} \qquad \delta=\delta(\Lb_v,\Av)\,.
\label{Densities}
\end{equation}
Equation~\eqref{dis} shows that the constitutive functions $\varphi$ and $\delta$ completely characterise the material response. The corresponding reduced constitutive functions will be introduced upon enforcing of the following invariance requests.

\section{Frame-indifference and structural frame-indifference}\label{SFISec}
The first requirement every physically grounded theory must obey to is the requirement of invariance under a change of frame. In the continuum mechanics community this is usually referred to as frame-indifference principle or simply as \emph{frame-indifference} \cite{Truesdell:1991,Gurtin:2010}. However, the Kr\"oner-Lee decomposition \eqref{kronerlee} may ask for an additional invariance requirement called \emph{structural frame-indifference}, which is the invariance under a change of frame in the structural space and is expressed by the trasformation laws~\eqref{F+} \cite{Green:1971,Gurtin:2010}. We assume both frame-indifference and structural-frame indifference as the starting points of our constitutive theory from which all the other requirements, including restrictions due to material symmetry and constraints, must follow.
\paragraph{Frame-indifference} Frame-indifference requires that the internal working $\Wc_i(\dot\ub,\Lb_v)$  be invariant under a change of frame defined by the transformation
\begin{equation}
	\Fb^\dagger=\Qb\Fb\,,\qquad \Qb\in \Rot
	\label{FrameInvariance}
\end{equation}
which maintains unaltered the viscous deformation $\Fv$ and, in view of \eqref{kronerlee}, induces the transformation $\Fe^\dagger= \Qb\Fe$ on the elastic deformation. It is a well-established results that \eqref{FrameInvariance} implies $\skw(\Sb\Fb^T )=\0$, i.e., the Cauchy stress $\Tb=J^{-1}\Sb\Fb^T$ is symmetric. Moreover, we stipulate the strain energy density to be frame-indifferent, that is 
\begin{equation}
	\varphi(\Qb\Fe,\Av) = \varphi(\Fe,\Av) \,\qquad \forall\;\Qb\in\Rot\,.
\end{equation}
This latter request is satisfied if and only if 
\[
\varphi=\phi(\Ce,\Av)\,,
\] 
that is, the strain energy density depends on the elastic deformation through the strain tensor $\Ce$, and on the orientation tensor $\Arem_v$. On noting that $\Av$ and $\Lb_v$ are unchanged by the transformation~\eqref{FrameInvariance},  the frame-indifference of the dissipation density $\delta$ readily follows.

\paragraph{Structural frame-indifference} The notion of structural frame-indifference leads to the consideration of the transformation laws \eqref{F+}, here rewritten as
\begin{equation}
\Fe^+= \Fe\Qb^T\,,\qquad \Fv^+=\Qb\Fv\,,\qquad \Qb\in\Rot\,,
\label{StructuralFrameIndifference}
\end{equation}
which keep unchanged the macroscopic deformation $\Fb=\Fe\Qb^T\Qb\Fv=\Fe\Fv$.
We shall say that the reduced constitutive function $\phi$ is structurally frame-indifferent if it is unchanged under \eqref{StructuralFrameIndifference}, which in turn leads to
\begin{equation}
\phi(\Qb\Ce\Qb^T,\Qb\Av\Qb^T) = \phi(\Ce,\Av)\,,\qquad \forall \Qb\in\Rot\,,
\label{SFIphi}
\end{equation}
since $\Av\mapsto\Qb\Av\Qb^T$.
Equation~\eqref{SFIphi} expresses the requirement of $\phi$ to be an isotropic function of $\Ce$ and $\Av$, or in other terms, that the material is transversely isotropic in its relaxed state, where the material symmetry axis is given by $\Av$ (see \cite{Zhang:1990}). Indeed, material symmetry and structural frame-indifference coincides in the present context.\\
The same invariance requirement \eqref{StructuralFrameIndifference} applied to the dissipation function $\delta$ yields
\begin{equation}
\delta(\Qb\Lb_v\Qb^T+\Omegab,\Qb\Arem_v\Qb^T) =\delta(\Lb_v,\Av)\,,\qquad \forall \Qb \in \Rot, \forall
 \Omegab\in \Skw\,,
 \label{SFIdelta1}
\end{equation}
in which the transformation \eqref{LvDv} of the velocity fields have been used.
Due to the arbitrariness of $\Qb$ and $\Omegab$ in \eqref{SFIdelta1}, one can choose $\Qb=\Ib$ and $\Omegab = -\Wb_v$ to show that $\delta(\Lb_v,\Arem_v)=\delta(\Db_v,\Arem_v)$, i.e., the function $\delta$ can only depend on the viscous stretching $\Db_v$ and not on the viscous spin $\Wb_v$.  In addition, by choosing $\Omegab=\0$, Eq.~\eqref{SFIdelta1} reduces to
\begin{equation}
\delta(\Qb\Db_v\Qb^T,\Qb\Av\Qb^T) = \delta (\Db_v,\Av)\,,\qquad \forall \Qb\in \Rot\,,
\label{SFIdelta2}
\end{equation}
which states that $\delta$ must be an isotropic function of $\Db_v$ and $\Av$, likewise the strain energy density $\phi$. One additional physical requirement on $\delta$ is the request of parity with respect to the argument $\Db_v$, that is to say, the request that the dissipation is unchanged by a change in the sign of the velocity, i.e., $\delta(-\Db_v,\Av)=\delta(\Db_v,\Av)$.

With the invariance requirements expressed by \eqref{SFIphi} and \eqref{SFIdelta2}, we are in the position of applying the representation theorem of isotropic functions to express $\delta$ and $\phi$ as polynomial functions of the scalar invariants of their arguments \cite{Liu:1982,Man:2018}. In particular, we consider the following integrity bases for $\phi$
 \begin{equation}
 I_1=\Ib\cdot\Cb_e\,,\; I_2=\Ib\cdot\Cb_e^\star\,,\;I_3=\det\Cb_e\,,\;I_4=\Arem_v\cdot\Cb_e\,,
I_5=\Arem_v\cdot\Cb_e^2\,,
\label{ElasticInvariants}
\end{equation}
with $\Cb_e^\star=J_e^2\Cb_e^{-T}$, thus, with a little abuse of notation, we write
\begin{equation}\label{recipe}
\phi(\Cb_e,\Arem_v)=\phi(I_1,I_2,I_3,I_4,I_5)\,.
\end{equation}
The same reasoning applied to the dissipation function $\delta$ lead us on introducing an analogous set of invariants like\footnote{The slightly different choice between the lists \eqref{ElasticInvariants} and \eqref{ViscousInvariants} gets the wish to recover the Ericksen theory of anisotropic viscous fluids within the present context.}
\begin{equation}
 J_1=\Ib\cdot\Db_v\,, J_2=\Ib\cdot\Db_v^2\,,J_3=\det\Db_v\,,J_4=\Arem_v\cdot\Db_v\,,J_5=\Arem_v\cdot\Db_v^2\,,
 \label{ViscousInvariants}
\end{equation}
which leads to the following form of $\delta$
\begin{equation}
\delta(\Db_v,\Av) = \delta(J_1,J_2,J_3,J_4,J_5)\,.
\label{deltaInvariants}
\end{equation}
With this form of $\delta$, the invariance of $\delta$ under a change in the sign of the velocity is satisfied if only even power of $J_1$, $J_3$ and $J_4$ appears in $\delta$. A separate form of the $\delta$ is often prescribed in the literature and so one is lead to introduce 5 constitutive coefficients, one for each invariant, which represent the independent characteristic times of the relaxation process. However, in order to be consistent with the Ericksen theory of anisotropic fluids a simpler choices will be made in the next section involving only 3 characteristic times.
%
%
%
%
%
%
\section{Dissipation inequality and evolution equations}
\label{vib}
%
We now write down the local form of the dissipation inequality \eqref{dis} by using the reduced form of the constitutive functions ${\phi}$ and $\delta$ as determined by structural frame-indifference. It reads as
 \begin{equation}
0\leq \varrho_r\, \delta(\Dv, \Av) = \Sb\cdot \dot{\Fb} +\sym\Gb \cdot \Db_v + \skw\Gb \cdot \Wb_v - \varrho_r\,\frac{\partial\phi}{\partial\Ce}\cdot \dot{\Ce} - \varrho_r \,\frac{\partial \phi}{\partial \Av}\cdot \dot{\Av}\,,
\label{dis2}
\end{equation}
and, by using the kinematic relationships \eqref{rates1} and \eqref{rates2}, one arrives at
\begin{eqnarray}
&&0\leq\varrho_r \,\delta(\Dv,\Av) = \big(\Sb \Fb^T-2 \varrho_r \,\Fe\frac{\partial \phi}{\partial \Ce}\Fe^T \big)\cdot \Db\nonumber\\
&+&\big( \skw\Gb+\varrho_r \big[ \Ce, \frac{\partial \phi}{\partial \Ce}\big]+ \varrho_r\big[\Av, \frac{\partial \phi}{\partial \Av}\big]\big) \cdot \Wb_v\label{dis3}\\
& +&\big( \sym\Gb + 2\, \varrho_r \,\sym(\Ce \frac{\partial \phi}{\partial \Ce}) - \varrho_r\,\sym((\Ib - \Av)\frac{\partial \phi}{\partial \Av}\Av) \big)\cdot \Dv\notag\,,
\end{eqnarray}
that must hold true for any tensors $\Db,\Wb_v,\Db_v$. The energy imbalance \eqref{dis3} can be used to derive, through the Coleman-Noll procedure \cite{Gurtin:2010}, thermodynamically consistent constitutive equations for $\Sb$ and  $\Gb$, granted the structurally frame-invariant constitutive representations \eqref{recipe} and \eqref{deltaInvariants} of $\varphi$ and $\delta$.

The first term in \eqref{dis3} yields the constitutive equation of the Piola-Kirchhoff stress $\Sb$ or, alternatively, of the symmetric Cauchy stress $\Tb=J^{-1}\Sb \Fb^T$, i.e.,
\begin{equation}
\Tb = 2\, \varrho_v\, J_e^{-1} \Fe \frac{\partial \phi}{\partial \Ce} \Fe^T\,,
\label{Cauchy}
\end{equation}
in which the  relationship $\varrho_v= J_e\,\varrho_r$ was used. The second term in \eqref{dis3} gives
\begin{equation}
\skw \Gb \cdot \Wb_v=\0\,,
\label{SecondTerm}
\end{equation}
since $\big[ \Ce, \frac{\partial \phi}{\partial \Ce}\big]+ \big[\Av, \frac{\partial \phi}{\partial \Av}\big]=\0$ (see Appendix) and $\delta$ is independent of $\Wb_v$. Previous equality must be satisfied for any $\Wb_v$ and so $\skw\Gb=\0$, that is the constitutively determined component of the inner action $\Gb$ is null.\footnote{The reactive part of this inner action does not enter the dissipation inequality, thus it is not restricted by it.}\\
It is worth remarking here that the multiplicative decomposition of $\Fb$ makes the viscous spin $\Wb_v$ indeterminate. In fact, structural frame-indifference implies the constitutive response to be independent of the transformation
\begin{equation}\label{trWv}
\Wb_v^+ = \Qb\Wb_v\Qb^T+\Omegab\,,\quad\forall \Qb\in \Rot\quad\textrm{and}\quad\forall\Omegab \in\Skw\,,
\end{equation}
such that $\Omegab=\dot{\Qb}\Qb^T$ and so one can arbitrary choose $\Wb^+$, e.g., if ${\dot{\Qb}=-\Qb\Wb_v}$, then $\Wb_v^+=\0$.\\ 
This indeterminacy has been longly debated in the field of large-strain elastoplasticity (see for instance \citep{Dafalias:1998} and \citep{Bruhns:2015}). In that framework,  in \citep{Gurtin:2005} it has been proved that, for isotropic materials, structural frame-indifference implies the inelastic spin to vanish, a result known in the literature as the \emph{Irrotationality Theorem} \citep{Gurtin:2010}. On the other hand, for anisotropic materials, structural frame-indifference only shows that the natural state can be determined within the equivalence class given by \eqref{StructuralFrameIndifference}. A choice commonly employed in the literature to overcome this indeterminacy is to choose $\Fb_v=\Fb_v^T$, such that $\Fv$ has only six independent components and their evolution is completely determined by the constitutive equation of the (symmetric) stretch rate $\Db_v$ \citep{Bruhns:2015}. Although legit, we prefer not to make this restricting choice on $\Fv$, but rather restrain its evolution by choosing
\begin{equation}
\Wb_v =\0\,,\label{ZeroSpin}
\end{equation}
and viewing it within the theory as an internal constraint on the motion of the body.\\ 
Finally, we identify in the third term the so-called Eshelby stress ${\Eb_{sh}=\Eb_{sh}(\Ce,\Arem_v)}$, defined by
\begin{equation}
\Eb_{sh}(\Ce,\Arem_v) =- 2 \,\varrho_r \,\sym(\Ce \frac{\partial \phi}{\partial \Ce}) + \varrho_r\,\sym((\Ib - \Av)\frac{\partial \phi}{\partial \Av}\Av)\,,
\label{Eshelby}
\end{equation}
and we rewrite the dissipation inequality as
\begin{equation}\label{formE}
\varrho_r\delta(\Dv,\Av) = \big(\sym\Gb - \Eb_{sh}\big)\cdot \Dv\,.
\end{equation}
On assuming that the inner action $\sym\Gb$ is made of elastic $\Eb_{sh}$ and dissipative $\Gb_{dis}$ parts, the difference $\sym\Gb-\Eb_{sh}$ in Eq.~\eqref{formE} defines the dissipative component $\Gb_{dis}$, which turns out to be restricted by the reduced dissipation inequality so to have
\begin{equation}\label{evG}
\varrho_r\delta(\Dv,\Av)=\Gb_{dis}\cdot\Db_v\geq 0\,.
\end{equation}
In accordance with \eqref{deltaInvariants}, we make the following assumption on the form of the dissipation density,
\begin{equation}
	\delta = \varrho_r^{-1}\mu\big(\tau_2\,J_2+\tau_4\,J_4^2 +2\tau_5\,J_5\big)\,,
	\label{dischoice}
\end{equation}
with $\mu$ the shear modulus of the material and $\tau_2$, $\tau_4$ and $\tau_5$ the \emph{characteristic relaxation times}, which control the viscous remodelling process. By using equations \eqref{evG} and \eqref{dischoice}, we obtain the dissipative component $\Gb_{dis}$ of the inner viscous stress,
\begin{equation}\label{gdis}
\Gb_{dis}= \mu(\tau_2\, \Dv + \tau_4 \,(\Dv \cdot\Av)\Av + \tau_5 \,(\Dv\Av+\Av\Dv))\,.
\end{equation}
\subsection{Evolution equations}
The (remodelling) evolution equations comes from the balance equations once the constitutive prescriptions derived through the dissipation inequality are used. In the present formulation, Eq.~\eqref{be}$_2$ has two uncoupled components: $\sym\Gb=\0$ and $\skw\Gb=\0$. The first component  determines the state of the body; indeed, by using equations \eqref{Eshelby}-\eqref{gdis} into the symmetric component of Eq.~\eqref{be}$_2$, we get the evolution equations of the viscous deformation: 
\begin{equation}
	\tau_2\,\Dv+\tau_4(\Dv\cdot\Av)\Av+\tau_5(\Dv\Av+\Av\Dv) = -\mu^{-1}\Eb_{sh}(\Ce,\Arem_v)\,.
	\label{EvEQ}
\end{equation}
Granted for Eq.~\eqref{SecondTerm} and its consequences, the second component is purely reactive and prescribe that the reactive part of $\skw\Gb$ is null.\\
Equation \eqref{EvEQ} is an evolution equation ($\Db_v$ is a strain rate) controlled by the three different relaxation times: $\tau_2$ is the contribution to the relaxation driven by the isotropic matrix, whereas $\tau_4$ and $\tau_5$ accounts for the additional dissipation associated to the material structure induced by the presence of fibres. If dissipation were uniquely caused by the matrix, that is, if we did not account for the dragged-by-viscous-relaxation fibres, we would have set $\tau_4=\tau_5=0$ in \eqref{EvEQ} and obtained the isotropic form of the  dissipation tensor used in \cite{Liu:2019}. On the other hand, if the viscous response were mainly due to the fibre contribution, as for instance in the experiments in \cite{Karduna:1997}, then $\tau_2=0$ and the viscous response would have been completely determined by $\tau_4$ and $\tau_5$. In this sense, the proposed form of the dissipation function can be seen as a generalization of the one used in \cite{Liu:2019}.\\
Moreover, the choice \eqref{dischoice} is also instrumental to obtain a constitutive equation of the Eshelby stress equal to the one introduced in \citep{Ericksen:1960} by Ericksen for anisotropic viscous fluids. Indeed, as shown in the next section, such a choice will make the model reducing to the one of anisotropic viscous fluids when the applied deformation is sufficiently slow. \\
\section{Asymptotic approximations: slow or fast deformations}\label{Asymptotics}
The objective of this section is to reconcile the developed model with the theory of hyperelasticity and viscosity. To do that, we study the asymptotic approximations of the model when visible deformations are \emph{fast} or \emph{slow}. We define the characteristic deformation time $\tau_d$ and introduce the dimensionless viscous stretching $\bar{\Db}$ as
\begin{equation}
\tau_d^{-1} = \vert \Db \vert, \qquad \bar{\Db}_v = \tau_d \,\Db_v\,,
\end{equation}
that has to be compared with the characteristic relaxation times in equation ~\eqref{EvEQ}, in order to study the asymptotic limits of the proposed model. Here and henceforth an overbar $\bar{\cdot}$ will be used to indicate the normalization with respect to $\tau_d$.

\paragraph{Slow deformations} We say that a deformation is \emph{slow} when the characteristic time $\tau_d$ is much longer that the relaxation times, which allows us to introduce the smallness parameter $\varepsilon$:
\begin{equation}
\textrm{max}\lbrace \tau_2,\tau_4,\tau_5 \rbrace = \tau_r \ll \tau_d \quad \Rightarrow \quad \varepsilon = \frac{\tau_r}{\tau_d}\ll 1\,.
\end{equation}
Then, we expand all the main kinematic variables in terms of $\varepsilon$ as follows:
\begin{equation}
	\Fe = \Ib + \varepsilon \Fb_1, \qquad \Fv = \big( \Ib - \varepsilon \Fb_1\big)\Fb+ o(\varepsilon)\,,
	\label{expasion1}
\end{equation}
and 
\begin{align}
&\Ce = \Ib + 2\,\varepsilon\,\Eb_e + o(\varepsilon)\quad\textrm{with}\quad \Eb_e = \sym \Fb_1\,,\\ 
	&\Av = \Ab + \varepsilon \,\big( 2\,(\Eb_e\cdot\Ab)\;\Ab-\Fb_1\Ab-\Ab\Fb_1^T\big)+o(\varepsilon)\,,\\
	&\bar{\mathbf{D}}_v= \bar \Db - \varepsilon \,\dot{\bar\Eb}_e+ \varepsilon\,\sym[\bar\Lb,\Fb_1]+o(\varepsilon)\,.
\end{align}
Accordingly, the first order approximation of the evolution equation \eqref{EvEQ} is
\begin{equation}
\eta_2 \,\varepsilon\, \bar{\Db} + \eta_4 \,\varepsilon(\bar{\Db}\cdot \Ab)\Ab+ \eta_5 \,\varepsilon\big( \Ab \bar{\Db} + \bar{\Db} \Ab\big) = \varepsilon\,\Eb_{sh}^{(1)}\,,
\label{1steveq}
\end{equation}
where it was used the fact that the Eshelby stress is expanded as $\Eb_{sh}=\Eb_{sh}^{(0)}+\varepsilon\,\Eb_{sh}^{(1)}$ and the zero-th order term, representing the stress at the relaxed state $\Ce=\Ib$ and $\Av=\Ab$,  vanishes. Indeed, by using the definition of the Eshelby stress~\eqref{Eshelby}, one can further prove that
\begin{equation}\label{Esh1}
\Eb_{sh} = -\varepsilon\, \Co[\Eb_e]+o(\varepsilon)\,,\quad\textrm{with}\quad
\mathbb{C}[\cdot]:=4\,\frac{\partial^2\phi}{\partial^2\Ce}\big\vert_{\Ce=\Ib,\Av=\Ab}\,
\end{equation}
that is the Eshebly coincides with the first order approximation of the Cauchy stress $\Tb = \varepsilon\, \Co[\Eb_e] +o(\varepsilon)$, which in turn gives through~\eqref{1steveq}
\begin{equation}
\Tb = \mu\,\tau_2\Db + \mu\,\tau_4 \big( \Db \cdot \Ab\big)\Ab + \mu \,\tau_5\big( \Db \Ab + \Ab \Db\big)\,, 
\end{equation}
the constitutive equation of the compressible Ericksen anisotropic fluid \cite{Ericksen:1960,Green:1964,Leslie:1966}.

\paragraph{Fast deformations} We say that a deformation is \emph{fast} when the characteristic time $\tau_d$ is much shorter that the relaxation times and so we introduce a smallness parameter $\varepsilon$ as
\begin{equation}
\textrm{min}\lbrace \tau_2,\tau_4,\tau_5 \rbrace = \tau_r \gg \tau_d \quad \Rightarrow \quad \varepsilon = \frac{\tau_d}{\tau_r}\ll 1\,,
\end{equation}
to expand the kinematical quantities\footnote{With a little abuse of notation, we now denote with $\Fb_1$ the first order approximation of the viscous deformation $\Fb_v$.}
\begin{equation}
\Fv = \Ib + \varepsilon \Fb_1, \qquad \Fe = \Fb \big(\Ib - \varepsilon \,\Fb_1\big)+o(\varepsilon)\,,
\end{equation}
and 
\begin{align}
&\Av = \Aref + \varepsilon\, \Ab_1	+o(\varepsilon)\\
&\Ce = \Cb - \varepsilon\,\Cb_1+o(\varepsilon)\\
&\bar{\Db}_v = \varepsilon\,\sym{\dot{\bar{\Fb}}_1} = \varepsilon\,\bar{\Db}_1\,.
\end{align}
In this case, the first order approximation of the evolution equation \eqref{EvEQ} is
\begin{equation}
	\eta_2 \varepsilon\bar\Db_1+\eta_4\, \varepsilon \big(\bar\Db_1\cdot \Aref\big)+\eta_5\varepsilon\big(\bar\Db_1\Aref+\Aref \bar\Db_1 \big) =\varepsilon\,\Eb_{sh}^{(0)}\,, 
\end{equation}
where $\Eb_{sh}^{(0)}$ represents  the Eshelby stress at the state $\Ce=\Cb$ and $\Av=\Aref$. Moreover,  the zero-order stress tensor is 
\begin{equation}
	\Tb = 2\,\varrho\,\Fb\frac{\partial \phi}{\partial\Ce}\big\vert_{\Ce=\Cb,\Av=\Aref}\Fb^T
	\label{1sthyper}
\end{equation}
which is, indeed, the constitutive equation of a transversely isotropic hyperelastic solid, whose direction of anisotropy is given by $\Aref$, the orientation tensor at the reference configuration. As such, Eq.~\eqref{1sthyper} shows that, under fast deformation, the model predicts a purely elastic response 
%
\section{Relaxation times and material response}
\label{examples}
%
With the aim of guiding the identification of the different constitutive coefficients of  the model,  we discuss some prototypical examples in which emphasis is put on the effects of the different relaxation times on the material response. We proceed by assuming that the elastic part of the material response is  the modified Holzapfel-Gasser-Ogden compressible model proposed in \cite{Nolan:2014}. In doing so, we assume that the elastic energy density $\phi$ has the following representation formula:
\begin{equation}
	\varrho_v\,\widehat\phi(I_1,I_3,I_4) = \frac \mu 2 \big(I_3^{-\frac 13}I_1-3\big)+\frac \mu 2\frac{\beta_1}{\beta_2}\big(\exp{\beta_2(I_4-1)^2}-1\big) + \frac \kappa 2 (I_3^{1/2}-1)^2\,,
\label{Nolan}
\end{equation}
which is the sum of isotropic, anisotropic and volumetric terms; $\mu$ and $\kappa$ are the shear and bulk moduli of the isotropic matrix, and $\beta_1$ and $\beta_2$ two positive coefficients weighting the reinforcement contribution of the fibres.\\ 

\subsection{Effects of relaxation time $\tau_2$}
We commence by investigating the effects on the material response of the relaxation time $\tau_2$, which controls the isotropic contribution in the dissipation and so can be interpreted as the relaxation time associated with the matrix material. To do so, we consider the longitudinal extension of a bar of length $L$, with coordinate $x\in[0,L]$, in which fibres are oriented along the longitudinal axis, i.e., $\theta_0=0$. The left-hand side of the bar is fixed, whereas force/displacement boundary conditions are applied at $x=L$. This example gives rise to a purely one-dimensional problem for which the (homogeneous) longitudinal stretch takes the form $\lambda(t)=\lambda_e(t)\lambda_v(t)$, with the corresponding viscous stretching $D_v(t)=\dot\lambda_v(t)/\lambda(t)$. Accordingly, the elastic energy density reduces to
\begin{equation}
\phi(\lambda_e)=\frac{\mu}{2} \frac{\beta _1}{\beta_2}\left(\exp\big(\beta _2 (\lambda_e^2-1)^2\big)-1\right)+\frac \kappa   2 (\lambda_e-1)^2+\frac \mu  2\big(\frac{\lambda_e^2+2}{\lambda_e^{2/3}}-3\big)
\label{1Denergy}
\end{equation}
and the (Cauchy) stress is given by $\sigma=\partial \phi/\partial \lambda_e$, that is
\begin{equation}
\sigma = 2 \beta _1 \mu  \lambda _e \left(\lambda _e^2-1\right) \exp\big(\beta _2 \left(\lambda _e^2-1\right)^2\big)+\kappa  \left(\lambda _e-1\right)+\frac{2 \mu  \left(\lambda _e^2-1\right)}{3 \lambda _e^{5/3}}\,.
\end{equation}
Here and henceforth, the dependence on time of stress and stretch is omitted for the sake of conciseness.

\begin{figure}
\begin{center}
\begin{tiny}
\def\svgwidth{1.\textwidth}
   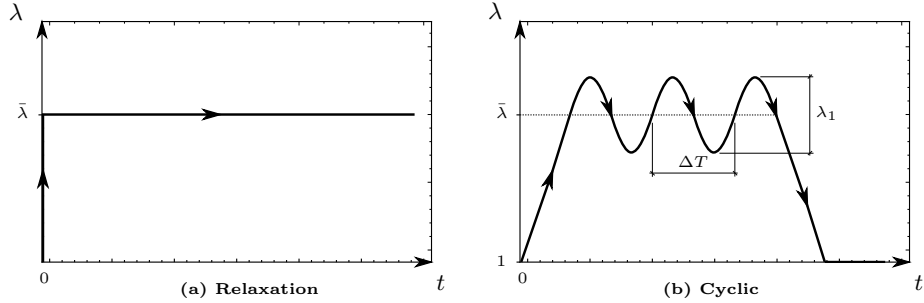
\end{tiny}
\end{center}
\caption{Time dependence of the externally imposed stretch $\lambda$ for relaxation (a) and cyclic (b) tests.}
\label{fig:2}
\end{figure}

The evolution of the viscous stretch $\lambda_v$ follows from the application of the energy imbalance \eqref{evG} and is written in terms of the visible stretch $\lambda$ as
\begin{align}
-&(\tau _2+\tau _4+2\, \tau _5)\, \lambda _v^3 \,\dot{\lambda} _v=\notag\\
&-\beta _1 \,\lambda ^2\,(\lambda ^2-\lambda _v^2)\; \exp\big(\beta _2 \big(\frac{\lambda ^2}{\lambda _v^2}-1\big)^2\big) - \frac{\kappa}{\mu} \lambda _v^3 (\lambda -\lambda _v)-\frac 23\,\lambda_v^2\,\big(\frac{\lambda_v}{\lambda}\big)^{\frac 23}(\lambda^2-\lambda_v^2)
\label{1Devolution}
\end{align}
to be solved with the initial condition $\lambda_v(0)=1$, once the dependence of the visible stretch $\lambda$ has been specified. In this respect, two different numerical experiments will be analysed representing relaxation and cyclic tests; the corresponding $\lambda$ is schematically depicted in Fig.~\ref{fig:2}.

In both experiments, displacement boundary conditions are applied at $x=L$, so to have the time dependence of the overall stretch as indicated in the figure. The following values of the constitutive parameters were assumed throughout this section
\begin{equation}
\mu=5~\text{MPa}\,,\qquad \kappa=20~\text{MPa}\,,\qquad \beta_1=4\,,\qquad \beta_2=0.5\,,
\label{constParam}
\end{equation}
unless otherwise noted. Concerning the relaxation times, Eq.~\eqref{1Devolution} shows that, in this one-dimensional setting, all relaxation times have the same effects on the dynamic evolution of the system, and so we fix $\tau_4=\tau_5=0$ and investigate the effects of $\tau_2$.
 
\begin{figure}
\begin{center}
\begin{tiny}
\def\svgwidth{1.\textwidth}
   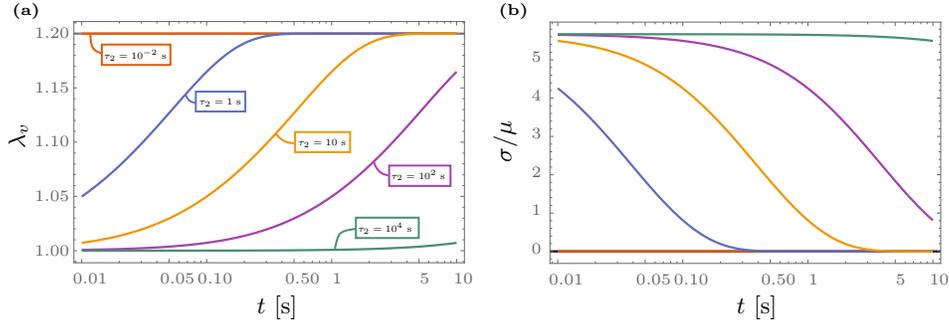
\end{tiny}
\end{center}
\caption{(a) Viscous stretch $\lambda_v$ and (b) dimensionless Cauchy stress $\sigma/\mu$ versus time $t$ in logarithmic scale for the one-dimensional relaxation test with  $\bar{\lambda}=1.2$, and $\tau_2=\lbrace 10^{-2},1,10, {10}^2, 10^4\rbrace$~s as per the color scheme in the left panel.}
\label{fig:3}
\end{figure}

During a relaxation test, the specimen initially at rest is suddenly (instantaneously) deformed and the stretch is thereafter maintained constant (Fig.~\ref{fig:2}, left). The results for $\bar{\lambda}=1.2$ and different values of $\tau_2$ are displayed in Fig.~\ref{fig:3}, where the whole range of viscoelastic effects is appreciated: from purely viscous, attained for long relaxation times, $\tau_2=10^4$~s, to purely elastic, for vanishing relaxation times, $\tau_2=10^{-2}$~s. In this latter case, in fact, the viscous stretch immediately reaches its asymptotic value $\bar{\lambda}=1.2$, meaning that $\lambda_e=1$ and the stress $\sigma$ is zero at all times, as one would expect from a viscous fluid. On the other hand, when the relaxation time is very long, e.g., $\tau_2=10^4$~s, the viscous stretch maintains its initial value $\lambda_v(0)=1$ and $\lambda_e=\lambda$, meaning that the stress immediately reaches its peak value as expected for an elastic solid. These results confirm the asymptotic analysis carried out in Sec.~\ref{Asymptotics}. 

\begin{figure}
\begin{center}
\begin{tiny}
\def\svgwidth{1.\textwidth}
   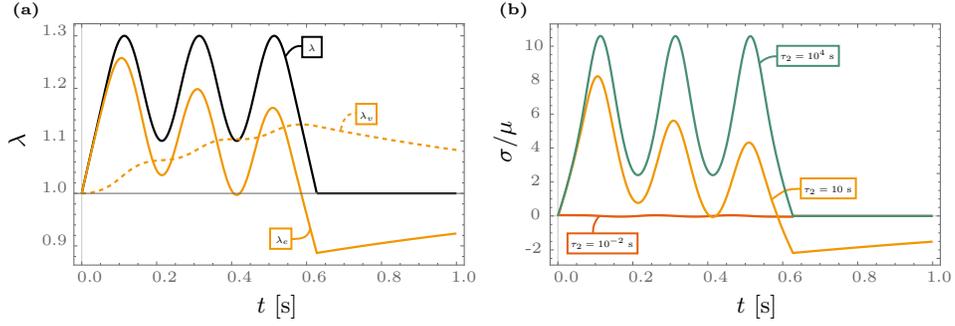
\end{tiny}
\end{center}
\caption{(a) Stretches $\lambda$, $\lambda_v$ and $\lambda_e(=\lambda/\lambda_v)$ and (b) dimensionless Cauchy stress $\sigma/\mu$ versus time $t$ for a one-dimensional cyclic test with  $\Delta T=0.2$~s. The left panel represents the solution of the evolution equation for $\tau_2=10$~s, whereas stresses in the right panel are calculated for $\tau_2=\lbrace 10^{-2},10, 10^4\rbrace$~s as indicated in the insets.}
\label{fig:4}
\end{figure}

\begin{figure}
\begin{center}
\begin{tiny}
\def\svgwidth{.5\textwidth}
   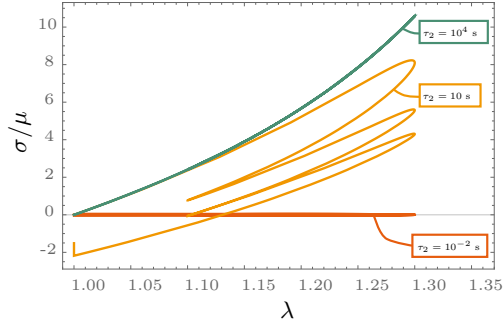
\end{tiny}
\end{center}
\caption{Dimensionless Cauchy stress $\sigma/\mu$ versus stretch $\lambda$ for the one-dimensional cyclic test with  $\bar\lambda=1.2$, $\Delta T=0.2$~s and $\tau_2=\lbrace 10^{-2},10,10^4\rbrace$~s as indicated in the insets.}
\label{fig:5}
\end{figure}

The same effects are observed in a cyclic test (Figs.~\ref{fig:4}-\ref{fig:5}). The externally imposed deformation is plotted in Fig.~\ref{fig:4}a with $\bar{\lambda}=1.2$, $\lambda_1=0.1$ and $\Delta T=0.2$~s (solid line), together with the corresponding viscous (dotted-yellow line) and elastic (continuous-yellow line) stretches obtained by solving the evolution equation \eqref{1Devolution} with $\tau_2=10$~s. The short time response, for  $t<0.05$~s, is purely elastic with $\lambda_v\approx 1$ and $\lambda_e\approx \lambda$. Thereafter $\lambda_v$ increases with a subsequent diminution of the elastic stretch that makes the stress decreases from its peak value (see the yellow curve in Fig.~\ref{fig:4}b); around $t=0.6$~s, when the externally imposed stretch goes back to $1$, the stress becomes negative meaning that the specimen would keep elongating due to the material flow, and a compressive force is needed to maintain the initial length. This negative force slowly decreases towards zero at times of the order of the characteristic time $\tau_2=10$~s, when the material flow interrupts. The effects of different relaxation times are apparently seen in Fig.~\ref{fig:4}b. For very short relaxation time, $\tau_2=10^{-2}$~s, the mechanical response is the one of a viscous fluid with viscosity $\simeq \mu \tau_2=0.05$~MPa$\cdot$s, and in fact the stress is almost zero in the figure. On the other hand, when the characteristic time is much longer than the time scale of the experiment, $\tau_2=10^{4}$~s, and at each time instant $\lambda_e\approx\lambda$ meaning that the stress attains the peak values dictated by the externally imposed stretch, resulting in a purely elastic behaviour. 
The stress-strain plots corresponding to the cyclic experiment are shown in Fig.~\ref{fig:5}. Again it is seen that, for the longest relaxation time, the response is purely elastic with the stress-strain curve having the characteristic strain hardening behaviour dictated by the exponential term in the elastic energy density \eqref{Nolan}, and a zero dissipation, seen by the vanishing area under the cycles. 
\subsection{Effects of relaxation times $\tau_4$ and $\tau_5$}
%
To investigate the effects of the relaxation times $\tau_4$ and $\tau_5$,  we consider a uniaxial confined stretch, in which the deformation is uniaxial and hampered in the plane orthogonal to the deformation axis. Typically, confined uniaxial stretches are implemented in the modelling of blood vessels undergoing large circumferential strain, with little or zero axial and radial strains \citep{Nolan:2014}. The following form of the visible deformation is  assumed
\begin{equation}
\Fb = \lambda\, \eb_1\otimes \eb_1 + \eb_2\otimes\eb_2+\eb_3\otimes\eb_3\,,
\label{confinedUniaxial}
\end{equation}
with $\{\eb_1,\eb_2,\eb_3\}$  a fixed orthonormal basis, and $\lambda$ an externally imposed stretch whose time dependence follows Fig.~\ref{fig:2}. To carry out a simple analysis which allows to discuss the role of $\tau_4$ and $\tau_5$,  we restrict our attention to a system in which fibres lie parallel to the $\eb_1$-axis. Under this circumstance, we assume that $\Fe$ and $\Fv$ share the same representation of $\Fb$, that is, they have the same principal directions, i.e.,
\begin{equation}
\Fe=\lambda_{e1}\,\eb_1\otimes \eb_1+\lambda_{e2}\,\eb_2\otimes \eb_2+\lambda_{e3}\,\eb_3\otimes \eb_3\,,
\label{Fe3D}
\end{equation}
and
\begin{equation}
\Fv=\lambda_{v1}\,\eb_1\otimes \eb_1+\lambda_{v2}\,\eb_2\otimes \eb_2+\lambda_{v3}\,\eb_3\otimes \eb_3\,.
\label{Fv3D}
\end{equation}
These expressions can be further simplified due the symmetry of the problem, for which $\lambda_{e3}=\lambda_{e2}$ and $\lambda_{v3}=\lambda_{v2}$, whereas the multiplicative decomposition of the deformation gradient \eqref{confinedUniaxial} leads to the following conditions on the internal stretches
\begin{equation}
\lambda_{e1}\lambda_{v1}=\lambda,\qquad \lambda_{e2}\lambda_{v2}=1.
\label{ConfinedStretches}
\end{equation}

In this three-dimensional setting, with equation \eqref{Nolan} on hand, one can use \eqref{Cauchy} to evaluate the Cauchy stress
\begin{equation}
\Tb = \kappa \,(I_3^{1/2}-1)\Ib + \mu\, I_3^{-5/6}(\Be-\frac 13 I_1\Ib)+2\,\mu\,\beta_1\, I_3^{-1/2}(I_4-1)\exp\big(\beta_2(I_4-1)^2\big)\Ab
\label{NolanCauchy}
\end{equation}
and the Eshelby stress
\begin{equation}
	\Eb_{sh} = -\kappa\,(I_3^{1/2}-1)\,\Ib -\mu \,I_3^{-1/3}\big(\Ce-\frac 13 I_1\,\Ib\big)-\mu\,\beta_1\, (I_4-1)\exp\big(\beta_2(I_4-1)^2\big)\Av\Ce\Av\,.
\end{equation}
Accordingly, the evolution equation \eqref{EvEQ} takes the form
\begin{align}
\tau_2\,\Dv&+\tau_4(\Dv\cdot\Av)\Av+\tau_5(\Dv\Av+\Av\Dv) = \nonumber\\
& \frac\kappa\mu\,(I_3^{1/2}-1)\,\Ib +I_3^{-1/3}\big(\Ce-\frac 13 I_1\,\Ib\big)+\beta_1\, (I_4-1)\exp\big(\beta_2(I_4-1)^2\big)\Av\Ce\Av\,,
\label{EvEqNolan}
\end{align}
to be solved with the initial condition $\Fv(0)=\Ib$.

\begin{figure}
\begin{center}
\begin{tiny}
\def\svgwidth{1.\textwidth}
   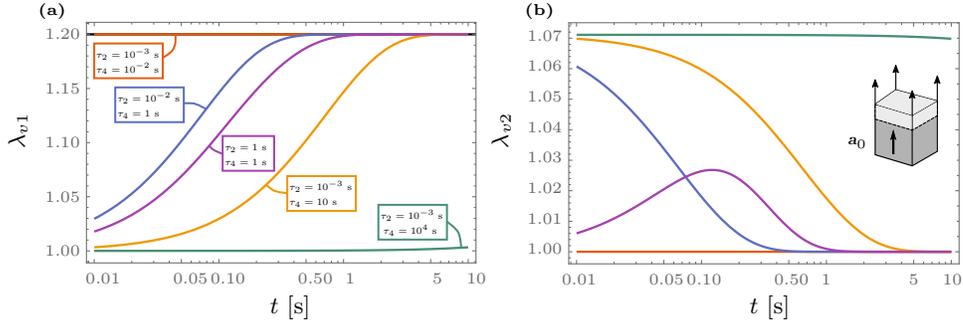
\end{tiny}
\end{center}
\caption{Longitudinal (a) and transverse (b) viscous stretches versus logarithmic time for a confined relaxation test with  $\bar{\lambda}=1.2$.}\label{fig:6}
\end{figure}

\begin{figure}
\begin{center}
\begin{tiny}
\def\svgwidth{1.\textwidth}
   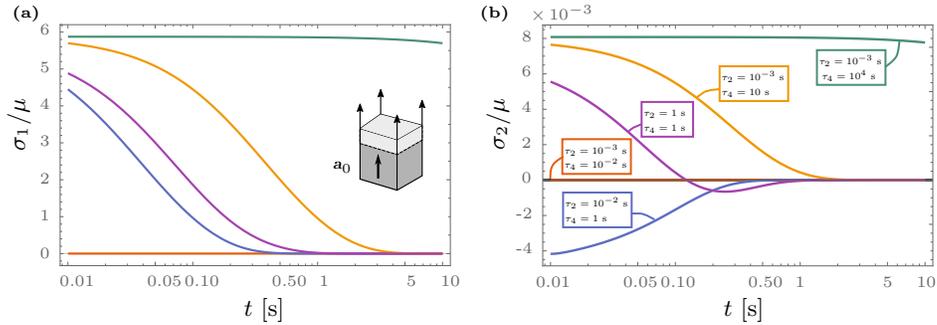
\end{tiny}
\end{center}
\caption{Longitudinal (a) and transverse (b) Cauchy stresses versus logarithmic time for a confined relaxation test with  $\bar{\lambda}=1.2$. The purple curve ($\tau_2=1$~s and $\tau_4=1$~s) is scaled by 0.01 for the sake of clearness}\label{fig:7}
\end{figure}

Equation~\eqref{EvEqNolan} together with \eqref{Fe3D} and \eqref{Fv3D} gives rise to two independent evolution equations, one in the longitudinal direction
\begin{align}
\mu  \, (\tau_2+\tau_4+2 \,\tau_5)\frac{\dot{\lambda}_{v1}}{\lambda_{v1}}&=\mu\,\beta_1 \,\lambda ^2\, (\lambda^2-\lambda_{v1}^2)\,\exp\!\Big(\beta_2 \big(\frac{\lambda ^2}{\lambda_{v1}^2}-1\big)^2\Big)\notag\\
&\kappa \, \Big(\frac{\lambda }{\lambda_{v1} \lambda_{v2}^2}-1\Big)+\frac 23\,\mu\,\lambda^{-\frac 23}\, \Big(\frac{\lambda ^2}{\lambda_{v1}^2}-\frac{1}{\lambda_{v2}^2}\Big)\big({\lambda_{v_1} \lambda_{v_2}^2}\big)^{\frac 23}\,,
\label{dyn1}
\end{align}
and one in the transverse direction
\begin{equation}
\mu \,\tau_2 \,\frac{\dot{\lambda}_{v2}}{\lambda_{v2}}=+\kappa \, \Big(\frac{\lambda }{\lambda_{v1} \lambda_{v2}^2}-1\Big)-\frac \mu 3\,\lambda^{-\frac 23}\, \Big(\frac{\lambda ^2}{\lambda_{v1}^2}-\frac{1}{\lambda_{v2}^2}\Big)\big({\lambda_{v_1} \lambda_{v_2}^2}\big)^{\frac 23}\,,
\label{dyn2}
\end{equation}
with $\lambda_{v1}(0)=1$ and $\lambda_{v2}(0)=1$. 
In addition, by substituting those expressions in \eqref{NolanCauchy}, one obtains the Cauchy stress in the longitudinal $\sigma_{1}$ and transverse $\sigma_{2}$ directions, that is
\begin{align}
\sigma_{1}&=2 \,\mu\,\beta_1 \,\frac{\lambda_{e_1}}{\lambda_{e_2}^2}\big(\lambda_{e_1}^2-1\big)\,\exp\!\Big(\beta_2 \big(\lambda_{e1}^2-1\big)^2\Big)+
\kappa  \big(\lambda_{e1} \lambda_{e2}^2-1\big)\,,\notag\\
&+\frac 23 \,\mu \, \big(\lambda_{e1}^2-\lambda_{e2}^2\big)\big(\lambda_{e1}\, \lambda_{e2}^2\big)^{-\frac 53}\\[0.2cm]
\sigma_{2}&=\kappa  \big(\lambda_{e1} \lambda_{e2}^2-1\big)-\frac \mu3 \, \big(\lambda_{e1}^2-\lambda_{e2}^2\big)\big(\lambda_{e1}\, \lambda_{e2}^2\big)^{-\frac 53}\,.
\end{align}
The results of the numerical simulations for relaxation and cyclic tests are shown in Fig.~\ref{fig:6}-\ref{fig:7}. The values of the constitutive parameters were those in \eqref{constParam}. Since $\tau_4$ and $\tau_5$ have the same effects on the dynamic evolution (see Eq.~\eqref{dyn1} and \eqref{dyn2}), we have set $\tau_5=1$~s and let $\tau_2$ and $\tau_4$ vary. 
The longitudinal and viscous stretches $\lambda_{v1}$ and $\lambda_{v2}$ in a relaxation test with $\bar{\lambda}=1.2$ are shown in Fig.~\ref{fig:6}. For all the curves, but the purple one, the characteristic time $\tau_2$ was set to $10^{-3}$~s, meaning that the isotropic contribution to the dissipation is negligible, whilst the characteristic times $\tau_4$ was changed as $\tau_4=\lbrace 10^{-2},1,10,10^4\rbrace$~s.
 The qualitative behaviour of the longitudinal viscous stretch for increasing characteristic times is similar to the one seen in the 1D example of Fig.~\ref{fig:3}a: a small characteristic time implies that the response is purely viscous and in fact $\lambda_{v1}$ immediately reaches the values of the externally imposed stretch $\bar\lambda$. On the other hand, if the relaxation time $\tau_4$ is very large, it is seen from \eqref{dyn1} than $\dot{\lambda_{v1}}\rightarrow 0$ and $\lambda_{v1}$ maintains its initial value, whereas, $\tau_2$ being very small causes $\lambda_{v2}$ to immediately reach its asymptotic value given by the right hand side of \eqref{dyn2}. Interestingly, a non-monotonic behaviour of the transverse stretch $\lambda_{v2}$ is observed when both the characteristic times $\tau_2$ and $\tau_4$ are equal to 1, which causes the transverse stress in Fig.~\ref{fig:7}b to be non-monotonic as well. This behaviour means that the specimen would contract in the transverse direction, if not laterally constrained, but then would start to expand when $\sigma_2$ becomes negative. For $\tau_2=10^{-2}$~s and $\tau_4=1$~s, $\sigma_2$ stays negative at all times, meaning that the apparent Poisson coefficient of the material is negative and if the lateral constraint were removed, the specimen would enlarge. The longitudinal stress $\sigma_1$ displays  the time decaying behaviour typical of relaxation tests.
\section{Finite element implementation}
\label{comp}
The equations of the model, listed in Tab.~\ref{tab:equations} and divided into balance, constitutive and evolution equations, are solved  in weak form through the finite element method implemented in the commercial software Comsol. 
\renewcommand*\arraystretch{1.25}
\begin{table*}
\centering
\caption{Recap of all modelling equations.}
\label{tab:equations}
\begin{footnotesize}
\begin{tabular}{llll}
\hline
balance 
& $\Div\Sb + \zb =\0\quad\text{in}\;\Bc_r\,,$ &$\ub=\ub\,\,\,\textrm{in}\,\,\partial_u\Bc_r \quad\textrm{and}\quad\Sb\,\nb=\mathbf{s}\,\,\,\textrm{on}\,\,\partial_t\Bc_r$ &\\
& $\Gb=\0\quad\text{in}\;\Bc_r$ &\\[0.2cm]
\multirow{2}{*}{constitutive}
&$\Sb=2\varrho_vJ_v^{-1}\Fe\frac{\partial \phi}{\partial \Ce}\Fv^{-T}$ &(first) Piola-Kirchhoff stress&\\
& \multicolumn{1}{l}{$\Eb_{sh} = - 2 \,\varrho_r \,\sym(\Ce \frac{\partial \phi}{\partial \Ce}) + \varrho_r\,\sym((\Ib - \Av)\frac{\partial \phi}{\partial \Av}\Av)$} &Eshelby stress&\\
&$\Gb_{dis}= \mu(\tau_2\, \Dv + \tau_4 \,(\Dv \cdot\Av)\Av + \tau_5 \,(\Dv\Av+\Av\Dv))$ &&\\[0.2cm]
\multirow{2}{*}{evolution}&\multicolumn{2}{l}{$\tau_2\,\Dv+\tau_4(\Dv\cdot\Av)\Av+\tau_5(\Dv\Av+\Av\Dv) = -\mu^{-1}\Eb_{sh}$}&\\
& \multicolumn{3}{l}{$\Wb_v=\0$}\\
&	&\\\hline
\end{tabular}
\end{footnotesize}
\end{table*}
Balance equations are written in weak form as
\begin{equation}\label{bala_weak1}
0= \int_{\Bc} \Sb\cdot\Grad\tilde\ub\,,\quad  0= \int_{\Bc}  \sym\Gb \cdot\tilde\Db_v\,,\quad 0= \int_{\Bc}  \skw\Gb\cdot\tilde\Wb_v\,,
\end{equation}
with $\Sb$ and $\sym\Gb=\Eb_{sh}+\Gb_{dis}$ constitutively assigned as specified in Tab.~\ref{tab:equations}, and $\skw\Gb$ identified by its reactive part  $\Rb_v$ necessary to maintain the constraint $\Wb_v=\0$.  On the other hand, the constraint is implemented in weak form as
\begin{equation}\label{bala_weak2}
0= \int_{\Bc} \Wb_v\cdot\tilde\Rb_v\,.
\end{equation}
%
%
\begin{figure}[h]
\begin{center}
\begin{tiny}
\def\svgwidth{1.\textwidth}
   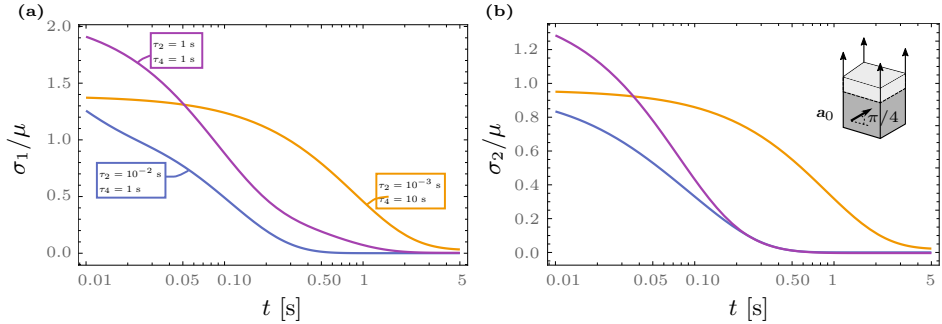
\end{tiny}
\end{center}
\caption{\label{fig:cf}  Longitudinal (a) and transverse (b) Cauchy stresses versus logarithmic time for a confined relaxation test with $\bar{\lambda}=1.2$ for fibres initially at $\theta_0=\pi/4$ (obtained with Comsol). }
\end{figure}
%
All in all, we solved the following problem: 
 find $(\ub,\Fb_v,\Rb_v)\in\Uc\times\Lin\times\Skw$ such that for all test fields $(\tilde\ub,\tilde\Fb_v,\tilde{\Rb}_v)$,
with $\tilde\Db_v=\sym(\tilde\Fb_v\Fb_v^{-1})$ and $\tilde\Wb_v=\skw(\tilde\Fb_v\Fb_v^{-1})$, equations \eqref{bala_weak1} and \eqref{bala_weak2} hold.  The first 3 equations deliver the standard balance of forces; the 6 equations \eqref{bala_weak1}$_2,3$, together with the 3 constraint equations \eqref{bala_weak2}, solve the viscous remodelling problem starting from an initial condition $\Fb_v(0)=\Ib$ in a \textit{fully coupled} form with the elastic problem \eqref{bala_weak1}$_1$.\\
As an example, we consider two problems corresponding to the relaxation of a confined and an unconfined longitudinal extension of a bar of length $L$, with coordinate $x\in[0,L]$. The problem was already solved in closed form in Sec.~\ref{examples} for fibres oriented at 0 and $\pi/2$. For the confined uniaxial extension, we consider fibres at $\pi/4$ in the $x-y$ plane, i.e., $\theta_0=\pi/4$, and we set $\ub=\0$ for the face at $x=0$ and $\ub=(\bar\lambda-1)\,\eb_1$ on the rest of the boundary. In the second case, fibres  were at $0$, $\pi/4$ and $\pi/2$ in the $x-y$ plane, and $\ub=\0$ on the face at $x=0$ and $\ub\cdot\eb_1=(\bar\lambda-1)$ on the rest of the boundary, so to leave unconstrained the transverse displacement $\ub-(\ub\cdot\eb_1)\,\eb_1$. In addition, we assumed $\ub(0)= (\bar\lambda-1)\,\eb_1$ everywhere at the initial time.\\
The integration algorithm used to solve the dynamic problem is the following:
\begin{enumerate}[a)]
\item \label{static} at time $t=0$, a \textit{static elastic problem} of uniaxial extension is solved (Eq.~\eqref{bala_weak1}$_1$) for $\bar\lambda$,  by assuming $\Fb_v=\Ib$ and hence $\Fb_e=\Fb$. The reference stress $\Sb$ is constitutively prescribed and corresponds to the Cauchy stress represented in the equation \eqref{NolanCauchy}: $\Sb=\Tb\Fb^\star$.
\item a \textit{time-dependent} analysis is carried out for the elasto-viscous relaxation problem corresponding to the extension $\bar\lambda$ (all Eqs.~\eqref{bala_weak1}),  and assuming as initial values of the variables those corresponding to the solution of the static problem in \ref{static}).
\end{enumerate}
%

%
Step a) allows the identification of the fibre re-orientation on the stress relaxation already evidenced in Fig.~\ref{fig:7}b for $\theta_0=0$. 
%
\begin{figure}[tb]
\begin{center}
\begin{tiny}
\def\svgwidth{.7\textwidth}
   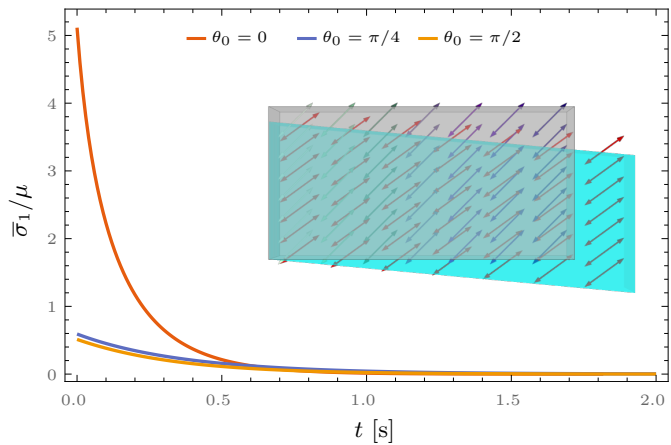
\end{tiny}
\end{center}
\caption{\label{fig:ncf}  Averaged stress $\sigma_1$ over the cross-section area for three different initial fibre angles $\theta_0=\lbrace 0,\pi/4,\pi/2\rbrace$. The inset shows a comparison between the reference and current configuration (at $t=2$~s) for fibres initially at $\theta_0=\pi/4$. All simulations were carried out with $\tau_2=1$~s and $\tau_4=1$~s.}
\end{figure}
%
In this case, longitudinal $\sigma_1$ and transverse $\sigma_2$ stresses are comparable and the non-monotonic pattern of $\sigma_2$ is much less evident, as Fig.~\ref{fig:cf} (left panel) shows; due to the fibre re-orientation, transverse stresses are indeed higher. 
Similar results hold for other choices of the characteristic times. Both longitudinal and transverse stresses  display the time decaying behaviour typical of relaxation tests.\\%
The second study focused on an unconfined longitudinal extension. In this case, fibre orientation strongly determines the deformed state of the body as Fig.~\ref{fig:ncf} shows. Therein, the reference configuration of the bar is shown grey coloured and the deformed configuration has been overimposed. Actually, the fibre reorientation is only slightly altered by the viscous deformations, yet the difference between the initial and final fibre orientation can be appreciated.

\section{Conclusions}

In this paper we have introduced a theoretical framework to define a nonlinear viscoelastic models compatible with the Ericksen theory of anisotropic fluids as well as the large strain theory of anisotropic hyperelastic solids.
The framework consists of a novel balance equation driving the passive material remodelling due to viscous deformation complemented with constitutive prescriptions based on three principles: the principle of indifference to change of observer, the principle of structural-frame indifference and the dissipation principle. The principle of structural frame-indifference yields a reduced form of the constitutive functions, namely the strain energy and dissipation densities, which turn out to be expressed in terms of scalar invariants of the elastic right Cauchy-Green strain tensor, $\Ce$, of the viscous rate of deformation $\Dv$ and of the orientation tensor $\bm A_v$ in the natural state. The dissipation principle lead us on identifying the inner remodelling action as the sum of an elastic Eshelbian component and a dissipative component whose representation form is completely determined by the choice of the dissipation function. Once the latter is introduced into the remodelling balance equation, the evolution equation of the natural state in terms of the viscous rate $\Dv$ is recovered. Moreover, to select among all the possible equivalent natural states and overcome the indeterminacy of the viscous spin $\Wb_v$, an internal constraint was introduced to prescribe the value of $\Wb_v$. 
There are some issues which have been left open and will addressed in future studies. \\
First, the proposed framework can be extended to incorporate the possibility for the fibre to reorient independently of the viscous deformation as proposed by the authors in \citep{Ciambella:2019,Ciambella2:2019}. The present model would then be recovered by imposing a suitable constraint on the fibre rotation. This extended framework would have interesting applications in the study of magneto-driven instabilities in fibre-reinforced structures \cite{Ciambella:2017_PRSA} and of bio-inspired morphing of fibres coated microcapsules \cite{Long:2015}.\\
Second, the model can be easily and promptly extended to encompass more complex dissipation functions dictated by the results of experiments in a purely phenomenological fashion as well as to encompass more complex elastic strain energies, also accounting for the \textit{so-called} equilibrium component, which would make the long-term stress not vanishing as seen in many biological tissues. 

\appendix
\section*{Derivation of Eq. \eqref{SecondTerm}}

The second term in Eq.\eqref{dis3} is here rewritten as
\begin{equation}
\big( \skw\Gb+\varrho_r \big[ \Ce, \frac{\partial \phi}{\partial \Ce}\big]+ \varrho_r\big[\Av, \frac{\partial \phi}{\partial \Av}\big]\big) \cdot \Wb_v\,,\end{equation}
and we now show that
\begin{equation}
\big[ \Ce, \frac{\partial \phi}{\partial \Ce}\big]+\big[\Av, \frac{\partial \phi}{\partial \Av}\big]=\0\,.
\label{toprove}
\end{equation}
By using the definition of the reduced dissipation function $ \varphi$, one has
\begin{equation}
\dfrac{\partial  \phi}{\partial \Ce}=\sum_{i=1}^5\phi_i\dfrac{\partial I_i}{\partial \Ce}
\end{equation}
with ${\phi}_i=\partial  \phi/\partial I_i$ and $I_1$, $I_2$, $I_3$, $I_4$ and $I_5$ the elastic invariants defined in Eq. \eqref{ElasticInvariants}. As such,
\begin{equation}
\frac{\partial I_1}{\partial \Ce}=\Ib,\quad \frac{\partial I_2}{\partial \Ce}=I_1 \Ib - \Ce,\quad \frac{\partial I_3}{\partial \Ce}=\Ce^\ast,\quad \frac{\partial I_4}{\partial \Ce}=\Av,\quad \frac{\partial I_5}{\partial\Ce}=\Ce\Av+\Av\Ce\,,\end{equation}
and the first term in \eqref{toprove} yields
\begin{equation}
\big[\Ce,\frac{\partial \phi}{\partial \Ce}\big]=\phi_4\big[\Ce,\Av\big] + \phi_5\big[\Ce,\Ce\Av+\Av\Ce\big]\,.
\label{primopezzo}
\end{equation}
For the second term, one has
\begin{equation}
	\frac{\partial \phi}{\partial \Av}=\phi_4\Ce+\phi_5 \big(\Ce\Av+\Av\Ce\big)
\end{equation}
since $\partial I_i/\partial \Av=0$ for $i=1,2,3$ and $\partial I_4/\partial\Av=\Ce$, $\partial I_5/\partial \Av = \Ce^2$. It hence reduces to
\begin{equation}
\big[\Av,\frac{\partial \phi}{\partial \Av}\big]=\phi_4\big[\Av,\Ce\big] + \phi_5\big[\Av,\Ce^2\big]\,.
\label{secondopezzo}
\end{equation}
On using the definition of the \emph{commutator} together with \eqref{primopezzo} and \eqref{secondopezzo}, one arrives at Eq. \eqref{toprove}.

\section*{Acknowledgments} 
This publication is based on the work supported by Sapienza Universit\`a di Roma under the project ''Mechanics of soft fibered active materials'' (No. RG11715C7CE2C1C4). PN  would like to thank MIUR (Italian Minister for Education, Research, and University) and the PRIN 2017, Mathematics of active materials: From mechanobiology to smart devices, project n. 2017KL4EF3, for financial support. JC acknowledges the support of MIUR through the project PRIN2017 n. 20177TTP3S.
\bibliographystyle{elsarticle-num} 
\bibliography{mybibfile}


%
%
%
\end{document}